\documentclass{article}
\usepackage{frascatiphys_R}

\begin{document}

\title{Rare Decays as Window to New Physics}
\author{
L.~M.~Sehgal\\ 
\em Institute for Theoretical Physics, RWTH Aachen \\ \em D-52056 Aachen, Germany
}
\maketitle
\baselineskip=11.6pt
\begin{abstract}
Rare decays of $K$ mesons are reviewed from the perspective of testing the ``ones'' and ``zeros'' of the standard model. Decays
$K^+ \to \pi^+ \nu \overline{\nu}$ and $K_L \to \pi^0 \nu \overline{\nu}$ probe the one-loop effective Hamiltonian for $s \to d \nu \overline{\nu}$,
and can constrain the $\rho, \, \eta$ coordinates of the unitarity triangle.
Decays such as $K_L \to \pi^0 l^+ l^-$, $K_L \to \mu^+ \mu^-$, $K^+ \to \pi^+ l^+ l^-$ and
$K_L \to \pi^+ \pi^- e^+ e^-$ involve short-distance effects, as well as
long-distance photon-induced contributions. Some comments are added on curious features of electroweak amplitudes in the
``gaugeless'' limit, and in the chiral electron limit $m_e \to 0$.
\end{abstract}

\baselineskip=14pt

\section{Ones and Zeros of the Standard Model}

The study of rare decays may be regarded as a part of the endeavour to test the principles of symmetry and symmetry-breaking underlying the standard
model of weak interactions. In any theory based on symmetries, the most important numbers are the ``ones'' and ``zeros'', the intensity
rules and selection rules. In the case of the standard model, the ones and zeros are associated with the unitarity of the quark-mixing
matrix, e.g.
\begin{eqnarray}
|V_{ud}|^2 + |V_{us}|^2 + |V_{ub}|^2 & = & 1 \ ({\rm ONE}) \label{BspOnes} \\
V_{ud} V_{us}^\ast + V_{cd} V_{cs}^\ast + V_{td} V_{ts}^\ast & = & 0 \ ({\rm ZERO}) \label{BspZeros}
\end{eqnarray}
Physically, Eq. (\ref{BspOnes}) expresses the universality of the lepton and hadron charged current couplings.
The present status of this relation may be judged from the empirical results\cite{PDG04} $|V_{ud}| = 0.9738 (5)$, 
$|V_{us}| = 0.2200 (26)$, and $|V_{ub}| = (3.67 (47)) \times 10^{-3}$, which satisfy Eq. (\ref{BspOnes}) to within a deficit 
$\Delta = 0.0033 (21)$.

The zero in Eq. (\ref{BspZeros}) represents a {\it unitarity triangle}, and is one of six that encode the structure of $CP$ violation
in the weak nonleptonic Hamiltonian. These triangles have diverse shapes, corresponding to the diversity of the elements $V_{ij}$.
There is, however, a unity in this diversity: all unitarity triangles have the same area $A_{\Delta}$, as a consequence of the fact that
$3 \times 3$ unitary matrices have an invariant property given by the Jarlskog parameter
\begin{equation}
J = {\rm Im} (\lambda_t \lambda_u^\ast) = {\rm Im} (\lambda_u \lambda_c^\ast)
\end{equation}
where $\lambda_u = V_{us} V_{ud}^\ast$, $\lambda_t = V_{ts} V_{td}^\ast$, $\lambda_c = V_{cs} V_{cd}^\ast$ with $\lambda_u + \lambda_c + \lambda_t = 0$,
and $|J| = 2 A_\Delta$. This invariant is a universal measure of $CP$ violation in weak phenomena. In addition, the existence of unitarity triangles
implies a unification of $CP$-violating and $CP$-conserving observables. The sides of a triangle are determined by the moduli $| V_{ij}|$, measurable
in $CP$-conserving processes. Knowledge of the sides fixes the angles, which are measures of $CP$
violation. This property, as well as the universal area of unitarity triangles, is a feature specific to a world with three generations.

The zero in Eq. (\ref{BspZeros}) has ramifications for flavour-changing neutral currents (FCNC). To order $G_F$, the weak neutral current has the structure
\begin{equation}
J_\mu^{\rm NC} = (\overline{d}, \, \overline{s}, \, \overline{b}) \gamma_\mu \frac{1 - \gamma_5}{2} V^\dagger V (d, \, s, \, b)^{\rm tr}
\end{equation}
and the unitarity of the matrix $V$ ensures the absence of non-diagonal terms. However, the symmetries which lead to the FCNC zero are broken in the
standard model by Yukawa couplings of the scalar doublet $( \varphi^+, \, \varphi^0)$ to fermions. For a typical doublet $(t, \, b)$, the Yukawa
interaction is 
\begin{equation}
{\cal L}_{\rm Y} = y_b (\overline{t}_L , \, \overline{b}_L) \left( \begin{array}{c} \varphi^+ \\ \varphi^0 \end{array} \right) b_R +
y_t (\overline{t}_L , \, \overline{b}_L) \left( \begin{array}{c} \varphi^{0 \dagger} \\ - \varphi^- \end{array} \right) t_R
\end{equation} 
with $y_b = \sqrt{2} m_b / v$, $y_t = \sqrt{2} m_t / v$ (note that $y_t$ is very nearly unity). These Yukawa couplings break
chiral symmetry and give rise to a FCNC interaction like $(\overline{s} d)_{V-A} (\overline{\nu} \nu)_{V-A}$ at the level of one-loop (box and penguin)
diagrams. Thus a typical FCNC amplitude has the form
\begin{equation}
A_{\rm FCNC} = G_F [ 0 ] +  G_F \alpha \sum_{i = u,c,t} \lambda_i f (m_i) .
\end{equation}

\section{Rare $K$ Decays}
\subsection{Golden Modes: $K^+ \to \pi^+ \nu \overline{\nu}$ and $K_L \to \pi^0 \nu \overline{\nu}$}

These two channels can be computed in an essentially model independent way from the effective Hamiltonian 
for $s \to d \nu \overline{\nu}$. The hadronic matrix element $\langle \pi^+ | (\overline{d} s)_{V-A} | K^+ \rangle$
can be related to the $K_{l3}$ matrix element, and long-distance effects are negligible\cite{Rein;Sehgal}.
The effective Hamiltonian derived from the box and penguin diagrams is\cite{Buras}
\begin{equation}
H_{\rm eff} = \frac{G_F}{\sqrt{2}} \frac{\alpha}{2 \pi \sin^2 \theta_W} \left[ \lambda_c X_{NL} + \lambda_t X_t \right] (\overline{d} s)_{V-A} 
(\overline{\nu} \nu)_{V-A} \label{Heffin21}
\end{equation}
where $X_{NL}$ is a small contribution due to $c$-quarks, and the dominant term is
\begin{equation}
X_t ( x_t) = \frac{x_t}{8} \left[ - \frac{2+x_t}{1-x_t} + \frac{3 x_t - 6}{(1-x_t)^2} \ln x_t \right]
\end{equation}
with $x_t = m^2_t/m^2_W$. In a limited domain of $m_t$, $X_t$ may be approximated as
\begin{equation}
X_t (x_t) = a + b x_t
\end{equation}
The dominant term in the effective Hamiltonian Eq. (\ref{Heffin21}) is then
\begin{equation}
H_{\rm eff} = \frac{G_F}{\sqrt{2}} \frac{\lambda_t}{4 \pi^2} \left[ \frac{1}{2} a g^2 + b y_t^2 \right] (\overline{d} s)_{V-A} (\overline{\nu} \nu)_{V-A}
\end{equation}
This expression reveals the two types of forces that are at work in FCNC decays: gauge forces associated with the gauge coupling $g \, (=e / \sin \theta_W)$
and Yukawa forces associated with the top-quark Yukawa coupling $y_t$. The latter force is independent of the gauge coupling, and exists even when
$g$ is switched off. It is the subtle interplay of these forces that one is testing in the study of FCNC processes.

\pagebreak

A thorough analysis of the decays $K^+ \to \pi^+ \nu \overline{\nu}$ and $K_L \to \pi^0 \nu \overline{\nu}$ has been carried out by Buras et al.\cite{Buras}.
The first reaction can be used to obtain $| V_{ts} V_{td}^\ast |$ and hence $[ (1-\rho)^2 + \eta^2 ]^{1/2}$, the second determines the $CP$-violating
parameter ${\rm Im} ( V_{ts} V_{td}^\ast) \sim \eta$. The two together can localise the $\rho, \, \eta$ coordinates of the unitarity triangle, and provide
a consistency check of the $(\rho, \, \eta)$ domain delineated by $B$-decays. The predicted branching ratios are
\begin{equation}
Br(K^+ \to \pi^+ \nu \overline{\nu}) = ( 7.8 \pm 1.2) \times 10^{-11},
\end{equation}
(to be compared with the experimental result $(14.7^{+13}_{-8.9}) \times 10^{-11}$ based on 3 events from the E949 and E787 experiments\cite{Redlinger}), and
\begin{equation}
Br(K_L \to \pi^0 \nu \overline{\nu}) = (3.0 \pm 0.6) \times 10^{-11}.
\end{equation}

\subsection{Decay Modes $K_L \to \pi^0 l^+ l^-$}

These decays receive contributions from three sources: (a) a $CP$-violating short-distance interaction
$s \to d l^+ l^-$, (b) a $CP$-conserving two-photon contribution associated with the decay
$K_L \to \pi^0 \gamma \gamma$, (c) an indirect $CP$-violating contribution associated with a one-photon transition $K_1 \to \pi^0 l^+ l^-$.
Accordingly, the decay amplitude has the structure
\begin{equation}
\begin{array}{ccccccc} A  & = & \underbrace{\eta \lambda^5 A_{sd}} & + & \underbrace{ \alpha^2 A_{2 \gamma}} & + & \underbrace{\alpha \epsilon A_{1 \gamma}}\\
& & ^{{\rm Direct} \, C\!P \!\!\!\!\! {\bf /}} & & ^{CP\!-\!{\rm conserving}} & &  ^{{\rm Indirect} \, C\!P \!\!\!\!\! {\bf /}} \end{array}
\end{equation}
The coefficients $\eta \lambda^5$, $\alpha^2$, $\alpha \epsilon$ have similar order of magnitude ($\eta \sim 0.3$, $\lambda \sim 0.2$, $\alpha \sim 10^{-2}$,
$\epsilon \sim 10^{-3}$). Data on the branching ratio and $\gamma \gamma$ spectrum of $K_L \to \pi^0 \gamma \gamma$ enable an estimate of $A_{2 \gamma}$.
The fact that the $2 \gamma$ state appears to be mainly $J=0$ implies that $A_{2 \gamma}$ is of importance mainly for the $K_L \to \pi^0 \mu^+ \mu^-$ channel.
The indirect $CP$-violating amplitude $A_{1 \gamma}$ is fixed (up to a model-dependent sign) by the observed branching ratio
for $K_S \to \pi^0 l^+ l^-$\cite{Batley}. A recent analysis obtains the prediction\cite{Isidorietal}
\begin{eqnarray}
Br(K_L \to \pi^0 e^+ e^-) & = & (3.7 \pm 1.0) \times 10^{-11} \\
Br(K_L \to \pi^0 \mu^+ \mu^-) & = & ( 1.5 \pm 0.3) \times 10^{-11} \nonumber
\end{eqnarray}

\subsection{Decay $K_L \to \mu^+ \mu^-$}

The decay $K_L \to \mu^+ \mu^-$ is subject to a unitarity bound associated with the $2 \gamma$ intermediate state\cite{Sehgal69}, given by
\pagebreak
\begin{equation}
R^K = \frac{\Gamma(K_L \to \mu^+ \mu^-)}{\Gamma(K_L \to \gamma \gamma)} \geq \frac{\alpha^2}{2 \beta} \frac{m^2_\mu}{m^2_K} \left( \ln \frac{1+ \beta}{1-\beta}
\right)^2 = 1.2 \times 10^{-5}
\end{equation}
where $\beta = ( 1- 4 m^2_\mu/ m^2_K )^{1/2}$. The measured value of $R^K$ is just $4 \%$ above the unitarity limit:
\begin{equation}
R^K_{\rm exp} = ( 1.238 \pm 0.024 ) \times 10^{-5}
\end{equation}
This excess can be interpreted as an estimate of the quantity
\begin{equation}
| A_{\rm disp}(2 \gamma) + A_{\rm s-d} |^2,
\end{equation}
where $A_{\rm disp}(2 \gamma)$ is the dispersive part of the $2 \gamma$ contribution, and $A_{\rm s-d}$ is the contribution of
the short-distance interaction $(\overline{s} d) (\overline{l} l)$. Such an analysis requires a model for the form factor of the two-photon vertex 
$K_L \to \gamma^\ast \gamma^\ast$\cite{Isidori}. In principle, access to the real and imaginary parts of
the $K_L \to \mu \overline{\mu}$ amplitude is also possible by studying the decay $K_L \to \mu^+ \mu^- \gamma$ in the soft-photon region where the
bremsstrahlung and Dalitz pair amplitudes for this process interfere\cite{Poulose;Sehgal}.

\subsection{Decay $K_L \to \pi^+ \pi^- e^+ e^-$}

The decay $K_L \to \pi^+ \pi^- e^+ e^-$ is calculable in terms of empirical knowledge of the radiative transition $K_L \to \pi^+ \pi^- \gamma$. It reveals a
remarkable $CP$-violating, $T$-odd asymmetry, which is triggered by the small $\epsilon$ impurity in the $K_L$ wave-function\cite{KpipieePapers}.

The $K_L \to \pi^+ \pi^- \gamma$ amplitude is the sum of a bremsstrahlung component, proportional to the $CP$-violating parameter $\eta_{+-}$,
and a direct $M1$ term obtained by a fit to the photon energy spectrum. The $e^+ e^-$ pair in $K_L \to \pi^+ \pi^- e^+ e^-$ is interpreted as an 
internal conversion of the photon in $K_L \to \pi^+ \pi^- \gamma$. The theoretical analysis leads to the prediction
\begin{equation}
\frac{d \Gamma}{d \phi} = \Gamma_1 \cos^2 \phi + \Gamma_2 \sin^2 \phi + \Gamma_3 \sin \phi \cos \phi
\end{equation}
where $\phi$ is the angle between the $\pi^+ \pi^-$ and $e^+ e^-$ planes. The last term is odd under $CP$ as well as $T$, and gives rise to an asymmetry
\begin{equation}
A_\phi = \left. \left( \int_0^{\pi/2} + \int_{\pi}^{3 \pi/2} - \int_{\pi/2}^\pi - \int_{3 \pi/2}^{2 \pi} \right) \frac{d\Gamma}{d\phi} d \phi \right/
\int_0^{2 \pi} \frac{d\Gamma}{d\phi} d \phi
\end{equation}
The predicted value was $14 \%$\cite{KpipieePapers}, and is in excellent agreement with the measured value\cite{Ledovskoy;Lai}
\pagebreak
\begin{equation}
A_\phi = \left\{ \begin{array}{cc} 13.7 \pm 1.4 \pm 1.5 \% & ( {\rm KTeV} ) \\ 14.2 \pm 3.6 \% & ({\rm NA48}  ) \end{array} \right.
\end{equation}
In addition, the distribution of the $\pi^+ \pi^-$ system in the final state confirms the presence of an $s$-wave amplitude, corresponding to a
mean-square $K^0$ charge radius
\begin{equation}
\langle R^2 \rangle_{K^0} = \left\{ \begin{array}{cc} -0.077 \pm 0.014 \, fm^2 & ( {\rm KTeV} ) \\ -0.09 \pm 0.02 \, fm^2 & ({\rm NA48}  ) \end{array} \right.
\end{equation}
in agreement with the theoretical expectation from vector meson dominance: 
$\langle R^2 \rangle_{K^0} = \frac{1}{2} \left[ \frac{1}{m^2_\phi} - \frac{1}{m^2_\rho} \right] = -0.07 \, fm^2$.

\subsection{Decays $K^+ \to \pi^+ e^+ e^-$ and $K_S \to \pi^0 e^+ e^-$}

These decays are determined mainly by the single photon intermediate state. The matrix elements have a similarity to that for the charged current decay 
$K^+ \to \pi^0 e^+ \nu$, and may be parametrised as
\begin{eqnarray}
{\cal A}(K^+ \to \pi^0 e^+ \nu) & = & \frac{G_F}{\sqrt{2}} \frac{f_+}{\sqrt{2}} \sin \theta_C (k+p)_\alpha \overline{\nu} \gamma^\alpha (1-\gamma_5) e \nonumber \\
{\cal A}(K^+ \to \pi^+ e^+ e^-) & = & a_+ \frac{G_F}{\sqrt{2}} \frac{\alpha}{\pi} f_+ \sin \theta_C (k+p)_\alpha \overline{e} \gamma^\alpha e \\
{\cal A}(K_S \to \pi^0 e^+ e^-) & = & a_S \frac{G_F}{\sqrt{2}} \frac{\alpha}{\pi} f_+ \sin \theta_C (k+p)_\alpha \overline{e} \gamma^\alpha e \nonumber
\end{eqnarray}

An early analysis\cite{Vainshtein} yielded the prediction $a_+ = -0.7$, $a_S = 2.4$. A simple model of $K^+ \to \pi^+ e^+ e^-$ relates the matrix element
to the weak two point vertex $K^+ - \pi^+$ and the charge radii of $K^+$ and $\pi^+$\cite{Burkhardt}. A similar model was used a long time ago\cite{Sehgal70}
to estimate the decay $K_S \to \pi^0 e^+ e^-$ in terms of the weak
vertex $K_2 - \pi^0$ and the charge radius of the $K^0$ meson. The $K^+ - \pi^+$ and $K_2 - \pi^0$ vertices are given by current algebra and PCAC:
\begin{equation}
\langle \pi^0 | H_w | K_2 \rangle = - \langle \pi^+ | H_w | K^+ \rangle = 2 F_\pi g
\end{equation}
where $g$ is the coupling constant for $K_1 \to \pi \pi$, and $F_\pi = m_N g_A/ g_{N N \pi} \approx 90 \, MeV$. With these values, the measured branching 
ratios of $K^+ \to \pi^+ e^+ e^-$ and $K_S \to \pi^0 l^+ l^-$ are well reproduced.

\section{Miscellaneous Remarks}

As noted above, the standard model contains gauge couplings $\{ g, \, g^\prime \}$, which
\pagebreak
conserve chirality, and 
Yukawa couplings $\{ y_f \}$ which are proportional to fer\-mi\-on masses and violate chirality.
It is the interplay of those couplings that determines the
strength of the FCNC interaction responsible for decays like $K^+ \to \pi^+ \nu \overline{\nu}$.

The reality of the Yukawa interaction as a force independent of gauge interactions is revealed if one considers the ``gaugeless" limit
of the standard model, viz. $g \to 0$ with $v = (\sqrt{2} G_F)^{-1/2}$ fixed. In this limit, studied by Bjorken\cite{Bjorken}, one has the remarkable consequence
that the electron is unstable, with decay width
\begin{equation}
\Gamma ( e^- \to \nu_e W^-) = \frac{\sqrt{2} G_F m_e^3}{16 \pi} = \frac{y_e^2}{32 \pi} m_e = ( 10.3 \, ns)^{-1} .
\end{equation}
Note that in the limit $g \to 0$, $m_W = gv/2 \to 0$. The electron decays purely by virtue of its Yukawa coupling $y_e = \sqrt{2} m_e / v$, and the
massless (longitudinal) $W$ it decays into is nothing but the massless Goldstone boson $\varphi^-$ of the scalar sector.

In a similar spirit, one can investigate the behaviour of amplitudes in the limit $y_e \to 0$ with $v$ fixed. A remarkable feature that emerges is that the
electron chirality is not conserved. This is
evident already at the level of QED: the cross section of helicity-flip Compton scattering is
\begin{equation}
\lim_{m_e \to 0} \sigma ( \gamma + e^-_L \to \gamma + e^-_R ) = 2 \pi \frac{\alpha^2}{s}
\end{equation}
Likewise, helicity-flip bremsstrahlung $e_L^- + {\cal N} \to e_R^- + {\cal N} + \gamma$ has the characteristic angular distribution\cite{Falk;Sehgal}
\begin{equation}
d \sigma_{hf} \sim \alpha \left( \frac{m_e}{E} \right)^2 \frac{d \theta^2}{\left( \theta^2 + \frac{m^2}{E^2} \right)^2}
\end{equation}
which, integrated over angles, gives a finite non-zero result in the limit $m_e \to 0$.

As a further interesting consequence\cite{Schulz;Sehgal} electrons in radiative muon decay
$\mu^- \to e^- \overline{\nu}_e \nu_\mu \gamma$ are not purely left-handed in the limit $m_e \to 0$. Despite the $V-A$ structure of the weak
interaction, there is a significant probability for electrons in $\mu$-decay to be right-handed. Such right-handed electrons are typically
accompanied by hard collinear photons. The contribution of these wrong-helicity electrons to the muon decay width is
$\Gamma_R = \frac{\alpha}{4 \pi} ( G_F^2 m_\mu^5 / 192 \pi^3 )$.

The above curiosities in the gaugeless limit or in the limit of a massless fermion may be of some relevance
when one contemplates the interplay of gauge couplings and Yukawa couplings in electroweak amplitudes.

\end{document}